\documentstyle[12pt,aasms]{article}
\eqsecnum
\begin{document}

\title{Does Dissipation in AGN Disks Couple to the Total Pressure?}

\author{Ethan T. Vishniac}

\affil{I: ethan@astro.as.utexas.edu, Department of Astronomy, University of
Texas, Austin, TX 78712}

\begin{abstract}
Recent work on the transport of angular momentum in accretion disks
suggests that the Velikhov-Chandrasekhar instability, in which a large
scale magnetic field generates small scale eddys in a shearing environment,
may be ultimately responsible for this process.  Although there is
considerable controversy about the origin and maintenance of this field
in accretion disks, it turns out that it is possible to argue, quite
generally, using scaling arguments, that this process is
sensitive to the total pressure
in an AGN disk, rather than the pressure contributed by gas alone.
We conclude that the resolution of the conceptual difficulties
implied by the presence of strong thermal and viscous instabilities
in radiation pressure and electron scattering dominated
does not lie in models that couple the total dissipation rate
to the gas pressure alone, or to some weighted mean of the gas
and radiation pressures.
\end{abstract}

\keywords{AGN}

\section{Introduction}

The conventional model for AGN is one in which matter is absorbed by a
supermassive black hole, typically with a mass of order $10^8 M_{\sun}$.
The matter flows inward towards the black hole event horizon through
a disk, which radiates away the heat generated by the dissipation of
orbital energy.  The dynamics of disk accretion are not well understood,
but the dissipation of energy is caused by the outward transport
of angular momentum.  Following Shakura and Sunyaev (1973) this transport
is modeled as the consequence of an effective viscosity $\nu\sim\alpha c_sH$,
where $H$ is the disk thickness and $c_s$ is the sound speed.
The mass accretion rate is taken to be close to the Eddington limit,
i.e. some fraction of a solar mass per year.  Given
this basic model the disk temperatures are typically
$10^5$ degrees or more.  The opacity of the disk is dominated
by electron scattering and its pressure is mostly due to radiation pressure.

Unfortunately, this model implies the presence of local instabilities so
rapid and powerful that the basic assumption of a steady state disk is
called into question (\cite{prp73}, \cite{le74}, \cite{ss76}).   Both thermal
instabilities, in which a small perturbation to the local thermal energy
density
grows, and viscous instabilities, in which a small perturbation to the local
surface density grows, are present on wavelengths almost as small as $H$.
Typical growth rates are  $\alpha\Omega$ for thermal instabilities and
$\alpha(H/r)^2\Omega$ for viscous instabilities.

Recently, Balbus and Hawley (\cite{bh91}, \cite{hb91}) have pointed out that
the
Velikhov-Chandrasekhar (VC) instability (which we have previously referred to
as the Magnetic Shearing Instability) provides a natural way to explain the
transport of angular momentum within an accretion disk.  This instability
(\cite{v59}, \cite{c61}) appears when a
magnetic field is present in a shearing flow.  The growth rate in
a keplerian disk is roughly
\begin{equation}
\Gamma_{VC}=3^{1/2} k_BV_A,
\end{equation}
where $k_B$ is the component of the wavevector of the perturbation in the
direction
of the magnetic field and $V_A$ is the Alfv\'en speed.
However, there is a minimum unstable wavelength $\sim V_A/\Omega$,
and the growth rate reaches a maximum ($\sim \Omega$) at slightly larger
scales.  This instability has the virtue that it necessarily transports angular
momentum in the right direction, i.e. outward.  In the strongly shearing
environment of an accretion disk we expect that the dominant role will be
played by the azimuthal component of the magnetic field.  We have previously
argued on theoretical grounds (\cite{vd92}) that in this case the
turbulent transport is dominated by the small, rapidly growing eddys.  This
argument
is apparently supported by preliminary numerical calculations (\cite{hb93}).
We expect that
the associated turbulent diffusivity is $\sim V_A^2/\Omega$, with an
associated value of $\alpha=(V_A/c_s)^2$.  If instead one
considers axisymmetric modes, which necessarily couple only to $B_z$, one
finds that the dominant eddys are large scale (or order $H$) with an associated
turbulent diffusivity of $V_AH$, where the Alfv\'en speed here is
$B_z/(4\pi\rho)^{1/2}$ rather than $B/(4\pi\rho)^{1/2}$ (\cite{hb92},
\cite{zdv93}),
and an $\alpha$ of $V_A/c_s$. In real disks such modes should be
suppressed by the more violent, smaller eddys associated with the azimuthal
field,
but we will see that our conclusions are insensitive to this point.
A major controversy regarding the role of the VC instability
is whether or not the motions induced by the turbulence are sufficient to
drive the magnetic field to a large amplitude, as suggested by Balbus and
Hawley (1991),
or whether this instability plays a purely dissipative role which must be
balanced
by a dynamo effect arising from unrelated fluid motions (e.g. \cite{vd92}).
We will see that our conclusions are insensitive to the resolution of this
controversy as well.

In this note we will argue that the VC instability in an AGN disk is
strongly coupled to both matter and radiation, so that its dissipative
effects should scale with the total (mostly radiation) pressure.
We also point out, contrary to previous arguments, that magnetic
buoyancy does not quickly eliminate any field whose saturation
Alfv\'en speed is a small fraction of the local sound speed.  Both of these
results are insensitive to the current controversy concerning the
maintenance of magnetic fields in disks.

\section{The Velikhov-Chandrasekhar Instability in an AGN Disk}

How can we tell if the VC instability couples to the radiation
pressure?  Clearly if the motions of the magnetic field and the
charged particles tied to it have no significant dynamical
interactions with the photons on the relevant length and
time scales then we can ignore the photons in applying the
instability to angular momentum transport.  The relevant
equipartition energy density for the magnetic field would
then be the thermal energy density of the gas.  This suggests
a series of questions.  First, can photons free stream through
the eddys created by the VC instability?  Second, can photons
damp the smallest VC eddys, thereby shifting the dynamics to
a larger scale, or blocking the Balbus-Hawley mechanism altogether?
Third, can a photon contained within an eddy escape in
one eddy turn-over time or less?  We will see that the answers to
these questions are "No'', ``No'', and ``Yes, but it doesn't
much matter''.

First, we consider whether or not photons
can free stream through the eddys created by the VC instability.
Let's consider the ratio of the photon
mean free path, $\lambda_{mfp}$, to the typical eddy size,
$V_A/\Omega$, where $V_A$ is the Alfv\'en speed and $\Omega$
is the local rotational velocity.  As mentioned above,
it has been claimed that the important eddys
are much larger, of order the disk thickness $H$, but clearly
if $\lambda_{mfp}<V_A/\Omega$ then the same conclusion must also
follow for larger eddys.  We have
\begin{equation}
{\Omega \lambda_{mfp}\over V_A}\sim {\Omega\over V_A\sigma_T n_e},
\end{equation}
where $n_e$ is the electron density.  However, for a disk supported
by radiation pressure and dominated by electron scattering the
vertical height can be obtained by balancing the average gravity
$\sim H\Omega^2$, with the acceleration per particle due to the radiative flux
$\dot M\Omega^2$.  We find that
\begin{equation}
H\sim {\dot M\sigma_T\over\mu c},
\end{equation}
where $\mu$ is the mean mass per electron.  It follows that
\begin{equation}
{\Omega \lambda_{mfp}\over V_A}\sim {\Omega\dot M\over \rho cH V_A}.
\end{equation}
Now since
\begin{equation}
\dot M\sim {\alpha\Sigma c_s^2\over\Omega}
\end{equation}
we have
\begin{equation}
{\Omega \lambda_{mfp}\over V_A}\sim \left({c_s^2\over c V_A}\right)\alpha
\end{equation}
Assuming that small scale eddys dominate the angular momentum
transport $\alpha\sim (V_A/c_s)^2$ and this ratio is $\sim V_A/c$, i.e.
small.  If instead we had assumed that the angular momentum
transport was dominated by eddys on a scale $\sim H$, then
we would have $\alpha\sim V_A/c_s$, but the relevant length scale
for the eddys would have been $H$ instead of $V_A/\Omega$ so the
final estimate of the ratio would have been unchanged.  If we
had taken a smaller estimate of $\alpha$, consistent with the
notion that radiation pressure doesn't contribute to the rate
of angular momentum transport, our conclusion would only have
been strengthened.  Photons {\it do not} free stream through VC
induced eddys in an accretion disk.

Second, we can ask if photons damp the smallest VC eddys, thereby shifting the
dynamics to a larger scale, or blocking the Balbus-Hawley mechanism altogether?
Now the appropriate ratio is the radiative damping rate to the eddy
turnover time.  Assuming the relevant eddys are small scale twists
in the large scale azimuthal field then the eddy turnover time is
just $\sim\Omega^{-1}$ and
\begin{equation}
{\tau_{eddy}\over\tau_{damping}}\sim \left({\Omega\over V_A}\right)^2
c\lambda_{mfp}
{\rho_{\gamma}\over\rho}\Omega^{-1}.
\end{equation}
This can be rewritten as
\begin{equation}
{\tau_{eddy}\over\tau_{damping}}\sim \left({c_s\over
V_A}\right)^2\left({P_\gamma\over P}\right)
{\Omega\dot M\over \rho c^2 H},
\end{equation}
or using the usual relationship between $H$ and $\dot M$
\begin{equation}
{\tau_{eddy}\over\tau_{damping}}\sim \left({c_s\over
V_A}\right)^2\left({P_\gamma\over P}\right)
\left({c_s\over c}\right)^2\alpha.
\end{equation}
Since $\alpha\sim (V_A/c_s)^2$ and $P_\gamma\sim P$ this is just $(c_s/c)^2$.
Again, a smaller value of $\alpha$ would only strengthen our conclusions.
If we had considered large scale, axisymmetric eddys, then
the eddy turnover time would have been longer, i.e. $H/V_A$, the eddy
length scale would have been longer, i.e. $H$, and the value of $\alpha$
would have been larger, i.e. $V_A/c_s$.  Our conclusion would be
unchanged.  The shear viscosity due to photons has a negligible effect
on the VC instability in AGN disks.

Finally, can a photon contained within an eddy escape in an
eddy turn-over time or less?  This is not equivalent to the preceding
question because such photons carry away a share of the eddy momentum
which is proportional to $\rho_\gamma/\rho\sim c_s^2/c^2$, which is small.
In fact, going through the argument in the preceding paragraph, but
omitting the factor of $\rho_\gamma/\rho$, we get that a typical
photon can just diffuse out of an eddy in one eddy turn-over time.
This means that the bulk viscosity of the fluid is not completely
negligible and the thermal conductivity of the fluid is high enough
that fluid in an eddy remains roughly at a constant temperature.
Nevertheless, the VC instability is a low frequency,
incompressible mode whose behavior is not sensitive to the
bulk viscosity or the thermal conductivity.
In particular we note that the fraction
of the mode energy which is contained in pressure fluctuations
is of order $(V_A/c_s)^2$, the remainder residing in kinetic
energy and magnetic field fluctuations.  Since only this fraction
of the mode energy can be dissipated by the bulk viscosity in
one eddy turn over time (given our result that photon's diffuse from
an eddy on this time scale), it follows that this damping mechanism
can have an effect of order unity only when the $\alpha\sim 1$
and $V_A\sim c_s$.

\section{Magnetic Buoyancy and the Saturation of the Magnetic Field}

One final loophole is that the saturation mechanism for the magnetic
field strength might depend on some process which does discriminate
between gas pressure and radiation pressure.  The proposal by
Balbus and Hawley (1991) that the VC instability leads to a
runaway growth of the magnetic field, at a rate $\sim \Omega$,
until $V_A\rightarrow c_s$ does not seem to allow for this possibility.
That is, if the local dynamics of the VC instability are responsible
for the growth and maintenance of the magnetic field then the fact
that this instability is insensitive to the distinction between
gas pressure and radiation pressure implies that the dissipation
of orbital energy in the disk is also insensitive to this distinction.

On the other hand, the theory that the magnetic field is driven by a
separate dynamo mechanism might allow for this possibility.
It becomes a question of whether or not the buoyancy of the field
depends in some critical way on whether the surrounding pressure is
due to radiation or gas.  For example, Sakimoto and Coroniti (1989)
have claimed that the high thermal conductivity of an AGN disk implies
that flux tubes within such a disk would be so buoyant, assuming
that dissipation couples to the total pressure, that the loss
of magnetic flux due to buoyancy would remove the disk's magnetic
field faster than it could conceivably be regenerated.  We will
now re\"examine this argument, bearing in mind the somewhat different
magnetic field dynamics implied by the VC instability.

We begin by reviewing the critical elements of their argument.
They modeled the buoyancy of small isolated flux tubes
within a disk and found that if the strength of the field was
allowed to scale with the local radiation pressure, then the
flux tubes were transparent and buoyancy losses were rapid.
The flux tubes were assumed to be almost axisymmetric
(that is, $B_r$ much less than $B_{\theta}$ and both constant
within a flux tube)  and to rise at
a speed limited only by their ability to adjust their internal
temperature to match their environment, and by the drag on the
flux tube.  The latter quantity was set equal to
\begin{equation}
-\left({C_d\over 2\pi}\right)(\pi a L)\rho_ev^2{v\over|v|},
\end{equation}
which is appropriate for the drag on a solid tube of radius $a$
and length $L$ moving with a speed $v$ through an approximately inviscid
medium with a density $\rho_e$.  The constant $C_d$ was set equal to $0.1$
in their simulations.  Assuming that the thermal
conductivity of the fluid is large we can estimate the
rise velocity by balancing the drag with the buoyant force
\begin{equation}
\left({V_A\over c_s}\right)^2 z\Omega^2\sim {V_{flux tube}^2\over a}.
\end{equation}
In other words, magnetic flux will be ejected from the disk at
a rate of approximately
\begin{equation}
{V_{flux tube}\over H}\sim {V_A\over c_s}\Omega \left({a\over H}\right)^{1/2}.
\label{eq:leave}
\end{equation}
Whether this is fast or not depends somewhat on the size of the flux tube, but
unless $a\ll H$ a typical flux tube will be ejected from the disk at the
Alfv\'en speed.
In practice the ejection rate can
be somewhat higher if the flux tube has a nonzero $B_r$ (as is
necessary to obtain an angular momentum flux from the global
stress on the field lines).  In that case the magnetic field
pressure in the flux tube rises quickly in the course of the
ejection.  On the other hand, for a flux tube of fixed size
whose magnetic pressure scales with the gas pressure alone,
the thermal conductivity is sharply reduced and the magnetic
flux loss rate will be much smaller.  Sakimoto and Coroniti
used this difference to argue that it isn't possible to construct
self-consistent disk models whose angular momentum transport is
due to the shearing of the magnetic field lines and whose magnetic
field pressure is proportional to the total pressure.

The presence of the VC instability affects this argument in a number
of ways.  First, no global value of $B_r$ is necessarily implied
in a model in which angular momentum transport is due to magnetic
field stresses.  Instead the angular momentum transport is due to
transient ripples in the field lines on scales of $V_A/\Omega$.
The globally averaged value of $B_r$ depends on the nature of the
dynamo mechanism supporting the field, but $B_rB_\theta$ is not
a significant source of angular momentum transport.  Second,
the typical size of magnetic field structures, including flux
tubes, is certainly no larger than the scale of the VC induced
turbulent eddys, $\sim V_A/\Omega$.  For a weak magnetic field this
scale will be proportionately smaller.  As we have shown in the
previous section, this implies that whether the magnetic field
pressure scales with the total pressure or just the gas pressure
typical structures in the field are always fairly conductive
thermally.  In neither case is buoyancy greatly hindered by the
ability of the flux tubes to drop to temperatures much below the
surrounding fluid.  Together these two points imply that the distinction
suggested by Sakimoto and Coroniti between models with a coupling
between dissipation and radiation pressure, and those without,
is not a meaningful one.  Magnetic buoyancy is a problem
for both, or for neither.

In fact, it appears to be a problem for neither, a point which we
have addressed in previous papers (\cite{vd92}, \cite{vd93}).
It turns out that the ejection rate given in Eq. (\ref{eq:leave})
is a significant overestimate.
Briefly put, the fact that the field lines form ripples of size
$V_A/\Omega$ at a rate of $\Omega$ implies that a flux tube of
size $V_A/\Omega$ will entrain a shifting volume of surrounding
material of roughly equal size on a time scale of roughly $\Omega^{-1}$.
This implies a drag of approximately $V_{flux tube}\Omega$.
Equating this to the buoyancy force implies
\begin{equation}
\left({V_A\over c_s}\right)^2 z\Omega^2\approx V_{flux tube}\Omega,
\end{equation}
or a typical upward speed of $\sim V_A^2/c_s$.  This in turn implies
a magnetic flux loss rate of $\sim (V_A/c_s)^2\Omega$.  For a strong
field this is very rapid, and for a weak field this is arbitrarily
slow.  In any disk dynamo model balancing this rate with the dynamo
growth rate will give a saturation level for the magnetic field
(and consequently for $\alpha$).  Of course, we could have assumed
still smaller flux tubes, but then the turbulent entrainment would
have been even more effective.  Larger flux tubes will self-destruct
due to the VC instability.

The bottom line is that the ejection of magnetic flux from a disk
is not particularly efficient unless the Alfv\'en speed is comparable
to the local sound speed, or equivalently unless the magnetic field
energy density is a large fraction of the total pressure.  There is
no mechanism that would tend to pin the magnetic field pressure to
the gas pressure.
This in turn implies that the dissipation of orbital energy in
an AGN disk couples to the total pressure, rather than just
the gas pressure.

\section{Discussion}

We have shown that if the Balbus-Hawley mechanism, which relies on
the VC instability, is responsible for angular
momentum transport in accretion disks, then the associated
dissipation scales with the total disk pressure, rather than just
the gas component.  In itself this is a somewhat negative result.
Its implication is that the strong thermal and viscous instabilities
found in AGN disks, cannot be due to coupling to an inappropriate
pressure.  Evidently, the pathological behavior of such disks
needs another cure.  Exactly what that cure might be is beyond
the scope of this paper, but we will mention two possibilities.
First, the pathology might be real, and might drive the disk
to some extremely inhomogeneous and/or optically thin state.
This outcome is implied by the view that the VC instability is
a self-sustaining process.  Second, the angular momentum transport might not
be described by purely local processes.  In this context this
would mean that while the VC instability is a local process
responsible for energy dissipation and angular momentum
transport, that the magnetic field dynamo is dependent
on nonlocal effects.  The wave driven dynamo (\cite{vjd90}, \cite{vd92}) is
an example of such a process.  Its implications for thermal
and viscous instabilities in AGN disks are discussed in Vishniac \&
Diamond (1993).

\acknowledgements

This work was supported by NASA grant NAGW-2418.  I would like to
thank Julian Krolik for a useful discussion.

\end{document}